\documentclass[journal=aamick,manuscript=article, layout=twocolumn]{achemso}

\usepackage[version=3]{mhchem} % Formula subscripts using \ce{}
\usepackage{multirow}
\usepackage{array}
\usepackage{xcolor}
\usepackage{dirtytalk}
\usepackage{placeins}
\usepackage{multirow}
\usepackage{array}
\newcolumntype{C}[1]{>{\centering\arraybackslash}p{#1}}
\usepackage[numbers]{natbib}
\usepackage{bibunits}

\usepackage{etoc}
\usepackage{bm}
\newcommand{\cm}{cm$^{-1}$}
\author{Satyam Sahu}
\affiliation[JHI]
{J. Heyrovský Institute of Physical Chemistry, Czech Academy of Sciences, Dolejškova 2155/3, 182 23 Prague, Czech Republic}
\email{satyam.sahu@jh-inst.cas.cz}

\author{Arsalan Hashemi}
\affiliation[]{European Laboratory for Learning and Intelligent Systems (ELLIS) Institute Finland, Maarintie 8, 02150 Espoo, Finland}
\alsoaffiliation{Department of Technical Physics, University of Eastern Finland, P.O. Box 1627, FI-70211 Kuopio, Finland}
\email{arsalan.hashemi@uef.fi}

\author{Mahdi Ghorbani-Asl}
\affiliation[]{Institute of Ion Beam Physics and Materials Research, Helmholtz-Zentrum Dresden-Rossendorf, 01328 Dresden, Germany}

\author{János Koltai}
\affiliation[ELU]
{Department of Biological Physics, Eötvös Loránd University, Pázmány Péter sétány 1/A, Budapest 1117, Hungary}

\author{Jan Maňák}
\affiliation[FZU]{Department of Material Analysis, Institute of Physics of the Czech Academy of Sciences, 182 21 Prague, Czech Republic}

\author{Bing Wu}
\affiliation[UCT]{Department of Inorganic Chemistry, University of Chemistry and Technology Prague, Technická 5, 166 28 Prague, Czech Republic}

\author{Aljoscha Söll}
\affiliation[UCT]{Department of Inorganic Chemistry, University of Chemistry and Technology Prague, Technická 5, 166 28 Prague, Czech Republic}

\author{Zdeněk Sofer}
\affiliation[UCT]{Department of Inorganic Chemistry, University of Chemistry and Technology Prague, Technická 5, 166 28 Prague, Czech Republic}

\author{Mikko Karttunen}
\affiliation[]{European Laboratory for Learning and Intelligent Systems (ELLIS) Institute Finland, Maarintie 8, 02150 Espoo, Finland}
\alsoaffiliation{Department of Technical Physics, University of Eastern Finland, P.O. Box 1627, FI-70211 Kuopio, Finland}

\author{Arkady V. Krasheninnikov}
\affiliation[]{Institute of Ion Beam Physics and Materials Research, Helmholtz-Zentrum Dresden-Rossendorf, 01328 Dresden, Germany}

\author{Matěj Velický}
\affiliation[JHI]
{J. Heyrovský Institute of Physical Chemistry, Czech Academy of Sciences, Dolejškova 2155/3, 182 23 Prague, Czech Republic}
\author{Otakar Frank}
\affiliation[JHI]
{J. Heyrovský Institute of Physical Chemistry, Czech Academy of Sciences, Dolejškova 2155/3, 182 23 Prague, Czech Republic}
\email{otakar.frank@jh-inst.cas.cz}

\title{Robust phonon engineering and symmetry-selective lattice dynamics in CrSBr$_{1-x}$Cl$_{x}$}

\keywords{CrSBr, atomic substitution, phonons, local perturbation, anisotropy, density functional theory, nonlinear}

\begin{document}

%\begin{tocentry}

%Some journals require a graphical entry for the Table of Contents.
%This should be laid out ``print ready'' so that the sizing of the

%Inside the \texttt{tocentry} environment, the font used is Helvetica
%8\,pt, as required by \emph{Journal of the American Chemical

%The surrounding frame is 9\,cm by 3.5\,cm, which is the maximum
%permitted for  \emph{Journal of the American Chemical Society}
%graphical table of content entries. The box will not resize if the
%content is too big: instead it will overflow the edge of the box.

%This box and the associated title will always be printed on a
%separate page at the end of the document.

%\end{tocentry}

\begin{abstract}
Atomic substitution provides a controlled route to engineer lattice dynamics in low-symmetry two-dimensional materials. Here, by combining polarization-resolved Raman spectroscopy and first-principles calculations, we investigate the evolution of phonon characteristics in CrSBr$_{1-x}$Cl$_{x}$ ($0 \leq x \leq \sim 0.5$) upon partial substitution of Br with Cl atoms. Progressive Cl substitution of Br induces systematic shifts of parent CrSBr out-of-plane $A_\textrm{g}$ phonon modes and activates additional Raman features. These features persist across different polarization configurations and excitation energies, reflecting substitution-induced symmetry lowering and local lattice perturbations. Explicit supercell phonon calculations combined with Raman $\Gamma$-density-of-states simulations identify these features as symmetry-lowered descendants of parent modes arising from alloy disorder. Complementary strain-dependent calculations reveal that anisotropic lattice compression plays a key role in renormalizing Cr--S dominated phonons. Under near-resonant excitation, stimulated Raman scattering-like amplification remains observable with increasing Cl content, highlighting the resilience of anisotropic electron--phonon coupling in this system.
\end{abstract}

%\section{New Concepts}
%\textcolor{blue}{Needed for some RSC journals.}\\
%Stimulated Raman scattering (SRS) is a nonlinear optical phenomenon with potential for tunable light sources, but its realization in layered materials has so far been limited in scope and spectral flexibility. In this work, we demonstrate compositionally tunable SRS in the anisotropic van der Waals material CrSBr, where partial substitution of Br with Cl enables systematic control of Raman-active phonon modes. This compositional tuning allows us to access a broad SRS spectral window from 240 to 290 cm$^{-1}$. By leveraging the sensitivity of SRS to phonon frequency shifts, we present a pathway to develop fine-resolution Raman lasers based on solid-state materials. Our results introduce a new strategy for achieving spectrally precise, compositionally programmable Raman gain media, with implications for integrated photonics, optical sensing, and nonlinear optics.
%%%%%%%%%%%%%%%%%%%%%%%%%%%%%%%%%%%%%%%%%%%%%%%%%%%%%%%%
%%%%%%%%%%%%%%%%%%%%%%%%%%%%%%%%%%%%%%%%%%%%%%%%%%%%%%%%%%%%%%%%%%%%%%%%%%%%%%%%%%%%%%%%%%%%%%%%%%
\section{Introduction}
Two-dimensional (2D) materials and their bulk layered counterparts provide a unique platform for engineering lattice dynamics due to their reduced dimensionality, weak interlayer bonding, and strong coupling between structural, electronic, and optical degrees of freedom \cite{Novoselov2016, Wang2012, Klein2023}. In this context, phonons play a central role, governing not only thermal and mechanical properties but also light–matter interactions, symmetry-breaking phenomena, and nonlinear optical responses \cite{Klein2023, Balandin2011, Balandin2012, Basov2016, Pawbake2023, Sahu2025}. Developing reliable strategies to systematically and predictably tune the phonon energies and behavior, therefore, remains an important challenge in low-dimensional materials.\par
Chemical substitution has emerged as one of the most effective approaches for phonon engineering in 2D systems, enabling systematic modification of lattice parameters and local bonding environments without extensively altering the overall crystal framework \cite{Rajapakse2021, Yang2021, Varade2023}. In transition metal dichalcogenides, for example, controlled substitution has been widely used to tune electronic structure and vibrational properties in a continuous manner \cite{Rajapakse2021, Varade2023, Chen2014, Cianci2024, Pucko2022, Olkowska2025, Zhao2018, Karthikeyan2019}. Across these systems, composition-dependent Raman shifts and the activation of disorder-induced modes have provided key insights into how local structural perturbations manifest in macroscopic vibrational responses \cite{Chen2014, Cianci2024}. However, most prior studies have focused on high-symmetry hexagonal lattices, where degeneracies and selection rules constrain the extent of observable phonon evolution \cite{Chen2014, Cianci2024}.\par
Low-symmetry 2D materials offer a complementary and largely unexplored opportunity for phonon engineering. CrSBr, a van der Waals layered antiferromagnetic semiconductor, is particularly well suited for this purpose. It combines robust air stability with a strongly anisotropic orthorhombic crystal structure, giving rise to a rich set of Raman-active phonon modes with well-defined polarization dependence \cite{Sahu2025, Mondal2025, Telford2023, Torres2023}. This intrinsic anisotropy makes CrSBr highly sensitive to lattice perturbations, allowing subtle structural modifications to be directly tracked through changes in its vibrational spectrum \cite{Telford2023, Torres2023}.\par
Halogen substitution in CrSBr provides a clean and controlled route to perturb the lattice. Recent works have shown that partial replacement of Br by Cl leads to a systematic contraction of the crystal lattice, most prominently along the $a$- and $c$-crystallographic axes, while leaving the optical band gap largely unchanged \cite{Telford2023, Badola2026}. Such composition-induced strain and local disorder are expected to strongly influence phonon frequencies, symmetry, and scattering pathways. Understanding how these effects manifest in the Raman response of CrSBr, particularly across varying composition and polarization, has so far remained limited.\par
In this work, we demonstrate systematic and robust phonon engineering in the mixed-halide series of CrSBr$_{1-x}$Cl$_x$ ($0 \leq x \leq 0.5$) using polarization-resolved Raman spectroscopy combined with first-principles calculations. Raman measurements performed under near-resonant and far-resonant excitation conditions reveal reproducible composition-dependent mode shifts and the emergence of additional Raman features (P$_1$–P$_3$). These features persist across different polarization configurations and excitation regimes, providing direct evidence of substitution-induced symmetry lowering and strain-driven phonon renormalization.\par
To establish the microscopic origin of these observations, we employ two complementary theoretical approaches. Explicit supercell phonon calculations capture the effects of statistical disorder on phonon frequencies and spectral broadening, while the Raman $\Gamma$-density-of-states (RGDOS) simulations reproduce the composition-averaged spectra and their polarization dependence. Together, these methods consistently identify the newly observed Raman features as symmetry-lowered descendants of the parent $A_\mathrm{g}$ modes, demonstrating that the atomic substitution acts as a reliable tuning parameter for lattice dynamics. We further show that the engineered phonon landscape governs the nonlinear Raman response under near-resonant excitation, linking controlled structural disorder to enhanced optical amplification in a low-symmetry 2D system.
%%%%%%%%%%%%%%%%%%%%%%%%%%%%%%%%%%%%%%%%%%%%%%%%%%%%%%%%%%%%%%%%%%%%%%%%%%%%%%%%%%%%%%%%%%%%%
\section{Methodological background}
\subsection{Sample preparation and characterization}
Bulk CrSBr and Cl-substituted CrSBr crystals were grown using the chemical vapor transport method and their atomic composition characterized using energy dispersive X-ray spectroscopy, as detailed in the supplementary data (Sections S1 and S2). Thin flakes were obtained by mechanically exfoliating the crystals directly onto SiO$_2$ (285 nm)/Si substrates.\\
Raman measurements were performed using a WITEC Alpha 300R system (Witec Instruments, Germany) equipped with a grating of 1800 lines/mm. Lasers with 1.96 eV and 2.33 eV excitation energies were focused onto the sample through a 100$\times$ (NA = 0.9) objective in the backscattering geometry. For the polarization-resolved measurements, the sample was fixed to a glass slide and mounted on a homemade goniometer, which was then rotated in intervals of about 20{\textdegree}  while keeping the polarizer and analyzer fixed and parallel to each other.\\
All the experiments were conducted at room temperature (295 K) under ambient conditions, unless otherwise stated.\\
The Raman spectra were fitted using Lorentzian functions in Origin 2024 to extract peak positions, widths, and integrated intensities.\\

\subsection{Computational details}
Density functional theory (DFT) calculations were performed with the VASP \cite{Kresse1993, Kresse1996} code using projector augmented wave (PAW) potentials \cite{Blochl1994}. Exchange-correlation effects were treated with the Perdew-Burke-Ernzerhof (PBE) functional \cite{Perdew1996}, and long-range dispersion was included using the Grimme DFT-D3 \cite{Grimme2011} scheme to account for the van der Waals interactions between layers. The structures were relaxed until the change in total energy between electronic steps was below $10^{-7}$ eV and all residual atomic forces were lower than 10$^{-3}$ eV/\AA. A plane-wave kinetic-energy cutoff of 500 eV was employed, and Brillouin-zone sampling used meshes corresponding to a $k$-point spacing of 0.2 \AA$^{-1}$. The vibration modes of the lattice were computed with Phonopy \cite{Togo2023} using the finite displacement method. Harmonic force constants were evaluated in a 4 $\times$ 3 $\times$ 2 supercell constructed from the primitive unit cell, which contains two monolayers to correctly reproduce the antiferromagnetic ground state at zero temperature. The supercell size was chosen to ensure convergence of the phonon frequencies and to accommodate low defect concentrations.\\
To describe vibrational disorder in the mixed-halide system, we adopt a two-host framework in which CrSBr$_{1-x}$Cl$_{x}$ is treated as a statistical ensemble of locally Br- and Cl-coordinated environments. Within this approach, phonon eigenvectors and force constants are inherited from the pristine constituents, CrSBr and CrSCl, while mass contrast and configurational averaging are treated explicitly using large supercells. This enables an efficient reconstruction of composition-dependent Raman responses while preserving the symmetry constraints of the parent compounds as described previously \cite{Hashemi2019, Ghafari2025, Berger2024, Oliver2020, Kou2020} and detailed in the supplementary data (Section S3).\\
In addition to the phonon--dispersion--based analysis and RGDOS simulations described above, conventional phonon DFT calculations were performed to directly obtain Raman-active mode frequencies and spreads. The computational details of these calculations are provided in the supplementary data (Section S4).

%%%%%%%%%%%%%%%%%%%%%%%%%%%%%%%%%%%%%%%%%%%%%%%%%%%%%%%%%%%%%%%%%%%%%%%%%%%%%%%%%%%%%%%%%%%%%%%%%%%%%%%%%%%%%%%%%%%%%%%%%%%%%
\section{Results and discussion}
\subsection{Lattice dynamics of pristine and Cl-substituted CrSBr}
We begin by establishing the structural accuracy and reliability of our first-principles calculations, which form the foundation for the subsequent analysis of the lattice dynamics and Raman response. Table \ref{table:structural} summarizes the optimized structural parameters obtained from DFT.\par
Using the experimentally reported lattice constants for bulk CrSBr, $a = 3.50$ \AA, $b = 4.76$ \AA, and $c = 15.92$ \AA\ as reference values \cite{Telford2020}, our calculations accurately reproduce the equilibrium crystal structure. In particular, the overall agreement in lattice vectors confirms that the chosen computational framework reliably captures the structural characteristics of CrSBr, and, thus, also of CrSCl.\par
Experimental extrapolations for Cl substitution at the Br site indicate that the evolution from CrSBr to CrSCl is accompanied by a pronounced anisotropic lattice response: a compressive strain of approximately 7.4\% along the out-of-plane $c$-axis, while the in-plane $b$-axis remains nearly unchanged and the $a$-axis contracts by up to 3\% \cite{Telford2023}. This experimentally inferred trend is well reproduced by our calculations, which yield contractions of 7.8\% and 2.3\% along the $c$- and $a$-axes, respectively, with a negligible change along the $b$-axis. Both the experiment and theory confirm the anisotropic nature of the substitution-induced structural response.\par
Further validation of the optimized structure is provided by the experimentally reported Cr–S–Cr bond angle of approximately 95\textdegree~ \cite{Lopez2022}, which is in very good agreement with our relaxed geometry. Together with the accurate reproduction of the lattice parameters, this confirms that our computational approach reliably captures both global and local structural features.\par

%%%%%%%%%%%%%%%%%%%%%%%%%%%%%%%%%%%%%%%%%%%%%%%%%%%
\begin{table*}[h]
    \centering
    \begin{tabular}{cccccccc}
    \hline
    \rule{0pt}{10pt}     & a ($\textrm{\AA}$) & b ($\textrm{\AA}$) & c ($\textrm{\AA}$) & Cr -- S ($\textrm{\AA}$) & X -- Cr ($\textrm{\AA}$) & X -- Cr -- X (\textdegree) & Cr -- S -- Cr (\textdegree) \\
    \hline
      CrSBr & 3.51 & 4.72 & 15.68 & 2.38 & 2.50 & 89.3 & 95.4\\
      CrSCl & 3.43 & 4.74 & 14.45 & 2.38 & 2.35 & 83.5 & 92.6\\
      \hline
    \end{tabular}
    \caption{The calculated structural parameters comprise the lattice constants a, b, and c; the Cr--S and X--Cr bond lengths (X = Br, Cl); and the X--Cr--X and Cr--S--Cr bond angles.}
    \label{table:structural}
\end{table*}
%%%%%%%%%%%%%%%%%%%%%%%%%%%%%%%%%%%%%%%%%%%%%%%%%%%%%%%%%%%%%%
\begin{figure*}[htbp!]
    \centering
    \includegraphics[width=15cm]{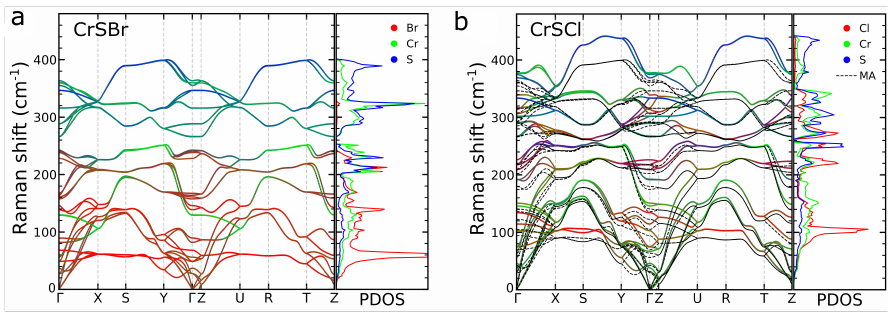}
    \caption{Phonon dispersion curves along the high-symmetry directions of the first Brillouin zone and the corresponding phonon density of states for bulk (a) CrSBr and (b) CrSCl. The phonon density of states is expressed in units of states/\cm. The high-symmetry points are defined as $\Gamma$(0, 0, 0), X($\frac{1}{2}$, 0, 0), S($\frac{1}{2}$, $\frac{1}{2}$, 0), Y(0, $\frac{1}{2}$, 0), Z(0, 0, $\frac{1}{2}$), U($\frac{1}{2}$, 0, $\frac{1}{2}$), R($\frac{1}{2}$, $\frac{1}{2}$, $\frac{1}{2}$), and T(0, $\frac{1}{2}$, $\frac{1}{2}$). Dashed lines in panel (b) show the dispersion calculated using mass approximation (MA).}
    \label{fig:ph_dispersion}
\end{figure*}
%%%%%%%%%%%%%%%%%%%%%%%%%%%%%%%%%%%%%%%%%%%%%%%%%%%%%%%%%%%%%%%%%%%%%
Having established a reliable computational framework for CrSBr, we now turn to its lattice dynamics, which form the basis for understanding the evolution of Raman-active phonons under halogen substitution. Bulk CrSBr crystallizes in the orthorhombic $D_{2h}$ structure with 12 atoms in the primitive unit cell. The symmetry classification of the corresponding zone-center phonon modes is well established in the literature \cite{Sahu2025, Mondal2025, Torres2023}; we therefore focus directly on the calculated phonon dispersions and their implications for lattice dynamics and Raman response.\par
Figure \ref{fig:ph_dispersion}a shows the calculated phonon dispersion of bulk CrSBr along high-symmetry directions of the first Brillouin zone. Using natural isotopic abundances, the average atomic masses of Cr, S, and Br are 51.99 u, 32.06 u, and 79.90 u, respectively. As expected from these mass differences, the low-frequency vibrational modes are dominated by Br vibrations, whereas higher-frequency branches primarily involve vibrations of the lighter Cr and S atoms. A narrow phonon gap of approximately 10 \cm~ separates the modes below 253.5 \cm~ and above 263.2 \cm. Below this gap, all atomic species contribute to the vibrational dynamics, while above it the Br contribution is strongly suppressed and the phonon modes are largely governed by Cr--S vibrations \cite{Torres2023, Xuan2023}.\par
For comparison, figure \ref{fig:ph_dispersion}b presents the phonon dispersion of the fully Cl-substituted analog, CrSCl. Owing to the substantially lower mass of Cl (35.45 u), halogen-dominated vibrational branches extend to higher frequencies relative to CrSBr, accompanied by a noticeable stiffening of the Cr--S related modes \cite{Xuan2023}. Importantly, no imaginary phonon frequencies are observed for either compound across the Brillouin zone, confirming the dynamical stability of pure CrSBr and CrSCl.\par
To isolate the role of the mass effects from changes in interatomic interactions, we also performed a model calculation in which the Br masses in CrSBr were artificially replaced by Cl masses while keeping the force constants unchanged. As shown by the dashed black curves in figure \ref{fig:ph_dispersion}b, this mass-only substitution fails to reproduce the full CrSCl phonon dispersion, with the most pronounced deviations occurring near the $\Gamma$ point and in the high-frequency branches. This behavior is expected, as the Cr--Cl bond is generally more ionic and therefore stiffer than the more covalent Cr--Br bond \cite{Telford2023}, leading to substantial force-constant renormalization upon halogen substitution.
%%%%%%%%%%%%%%%%%%%%%%%%%%%%%%%%%%%%%%%%%%%%%%%%%%%%%

\subsection{Composition-dependent Raman response}
%%%%%%%%%%%%%%%%%%%%%%%%%%%%%%%%%%%%%%%%%%%%%%%%%%%%%%%%%%%%%%%%
\begin{figure*}[h]
    \centering
    \includegraphics[width=15cm]{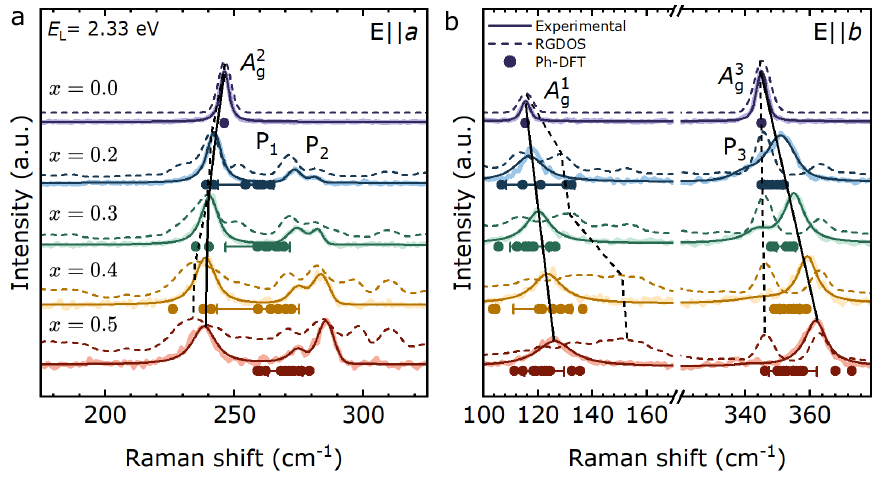}
    \caption{Raman spectra of Cl-substituted CrSBr measured under $E_\textrm{L}$ = 2.33 eV excitation at 100 {\textmu}W for laser polarization along the crystallographic $a$-axis (a) and $b$-axis (b), as a function of Cl concentration. Solid curves represent the fitted spectra, while the lighter shaded curves correspond to the raw experimental data. Dashed curves represent the RGDOS-simulated spectra, while solid circles underneath the spectra mark the phonon energies calculated using the phonon DFT for various Cl substitution configurations, with horizontal bars indicating the standard deviation arising from local configurational disorder. Vertical solid (experimental parent $A_\textrm{g}$ modes) and dashed (RGDOS-simulated parent $A_\textrm{g}$ modes) lines are guides to the eye.}
    \label{fig:Raman_spectra}
\end{figure*}
%%%%%%%%%%%%%%%%%%%%%%%%%%%%%%%%%%%%%%%%%%%%%%%%%%%%%%%%%%%%%%%%%%%%%%%%%%
We begin with the experimental Raman spectra of pristine CrSBr ($x = 0$). Under polarized excitation along the crystallographic $a$- and $b$-axes (solid curves in figure \ref{fig:Raman_spectra}a and \ref{fig:Raman_spectra}b, respectively), the spectra are dominated by the well-defined parent Raman modes at approximately 115 \cm~ ($A_\textrm{g}^1$), 246 \cm~ ($A_\textrm{g}^2$), and 345 \cm~ ($A_\textrm{g}^3$). Their frequencies, polarization dependence, and relative intensities are consistent with previously reported zone-center optical phonons in CrSBr and provide a clear reference for tracking the composition-induced changes \cite{Sahu2025, Mondal2025, Torres2023}.\par
Upon progressive Cl substitution, these parent modes exhibit systematic and mode-specific frequency shifts. The lowest-frequency $A_\textrm{g}^1$ mode undergoes a pronounced blue shift ($\sim 10$ \cm) with increasing Cl content, reaching 126 \cm~ at $x = 0.5$. This behavior reflects the progressive replacement of heavier Br atoms by lighter Cl atoms in the halogen-dominated vibrations of $A_\textrm{g}^1$. In contrast, the $A_\textrm{g}^2$ mode near 246 \cm, which primarily involves Cr--S displacements, shows a modest softening ($\sim 8$ \cm), while the high-frequency $A_\textrm{g}^3$ mode, which also involves Cr-S displacements, exhibits a monotonic blue shift ($\sim 17$ \cm), indicating stiffening of specific Cr-S stretching vibrations due to the change in the local environment. The persistence of these trends across the available composition range highlights the systematic and robust nature of the phonon evolution.\par
In addition to the shifts of the parent modes, new Raman features labeled P$_1$ -- P$_3$ emerge with increasing Cl concentration and become clearly discernible for $x \geq 0.2$. For excitation polarized along the $a$-axis, P$_1$ and P$_2$ appear near 273 \cm~ and 281 \cm, respectively, and shift to higher frequencies with increasing Cl content. For polarization along the $b$-axis, an additional feature P$_3$ emerges near 340 \cm~ and exhibits a similar blue-shift trend, as can also be seen in the supplementary data (Section S5).\par
To understand the microscopic origin of the composition-dependent Raman evolution, we combined explicit phonon calculations with Raman RGDOS simulations based on DFT. In the first approach, phonon frequencies were calculated for large supercells containing random Cl/Br distributions. The resulting discrete mode energies, shown as scattered points in figure \ref{fig:Raman_spectra}, represent the configurationally averaged shifts of the zone-center phonons. Their increasing spread with Cl substitution directly reflects the degree of local structural and vibrational disorder in the mixed-halide lattice. The similarity between these computed frequency distributions and the experimental linewidth broadening confirms that the observed spectral changes arise primarily from intrinsic disorder through mass contrast and strain fluctuations, rather than extrinsic defects.\par
To gain deeper insight into the Raman intensity and polarization behavior, we applied the RGDOS approach, which reconstructs the Raman response by projecting the supercell eigenmodes onto the Raman-active eigenvectors of the pristine constituents, CrSBr and CrSCl. This hybrid model captures the contrast in bond stiffness and mass between Br and Cl while retaining the correct lattice symmetry. The simulated spectra (shown as dashed lines in figure \ref{fig:Raman_spectra}) reproduce most of the key experimental trends starting with the upshift of halogen-dominated $A_\textrm{g}^1$ mode with an increase in the Cl content, while the $A_\textrm{g}^2$ mode redshifts along with broadening due to local strain and electrostatic perturbations.\par
A particularly significant outcome of the RGDOS analysis is the emergence of the additional Raman components within the $A_\textrm{g}^2$ and $A_\textrm{g}^3$ regions, corresponding to the experimentally observed P$_1$, P$_2$, and P$_3$ peaks. Mode-projection analysis confirms that P$_1$ and P$_2$ originate from local symmetry-lowered variants of the $A_\textrm{g}^2$ vibration, whereas P$_3$ arises from the perturbed $A_\textrm{g}^3$ manifold. In particular, the P$_1$ and P$_2$ modes are found to originate predominantly from halogen-substitution, reflecting the local chemical disorder within the mixed Br/Cl sublattice. By contrast, the P$_3$ mode is mainly driven by the Cr--S vibrations (Section S6). These symmetry-lowered local vibrations become Raman-active when the halogen sublattice loses long-range periodicity, giving rise to phonons with finite wavevector components that acquire Raman activity through disorder-induced symmetry breaking. The calculations also reproduce the increasing relative intensity of P$_1$ and P$_2$ with Cl concentration, reflecting the growing statistical weight of the mixed-halogen environment. Also, see Section S7 of the supplementary data for all the simulated compositions, i.e., CrSBr$_{1-x}$Cl$_{x}$, where $0 \leq x \leq 0.88$.\par
Overall, the RGDOS simulations reproduce the ordering and qualitative composition dependence of the Raman-active modes. However, for the Cr--S dominated $A_\textrm{g}^3$ mode, the calculations predict only a weak dependence on Cl substitution, in contrast to the clear and systematic upshift observed experimentally. This discrepancy indicates that the evolution of this mode cannot be fully described within a mass-only substitution picture using idealized bulk force constants and points to additional structural contributions that influence the vibrations.\par
In addition, the RGDOS simulations also predict several weak additional Raman-active components that are not resolved experimentally. These features could originate from low spectral-weight contributions of locally perturbed vibrational configurations and may be strongly broadened due to configurational averaging and disorder. A further source of the discrepancies stems from the different thermodynamic conditions: while the experiments were performed at room temperature, the simulations correspond to 0 K. Finite-temperature effects, including anharmonicity and thermal broadening, are therefore expected to further smear the weak features and renormalize phonon frequencies.\par
Importantly, when comparing our calculations with recent low-temperature Raman measurements \cite{Badola2026}, we find excellent agreement. In this regime, the calculated spectra reproduce the key experimental Raman features and enable a detailed assignment of the observed scattering response, indicating that the underlying lattice-dynamical description is robust and that much of the residual mismatch at room temperature can be attributed to thermal and disorder-related effects rather than fundamental shortcomings of the model.

%%%%%%%%%%%%%%%%%%%%%%%%%%%%%%%%%%%%%%%%%%%%%%%%%%%%%%%%
\subsection{Strain effects and limitations of the two-host model}
%%%%%%%%%%%%%%%%%%%%%%%%%%%%%%%%%%%%%%%%%%%%%%%%%%%%%%%%%%%%%%%%%%%%%%%%%%%%%%%%%%
\begin{figure*}[h]
    \centering
    \includegraphics[width=15cm]{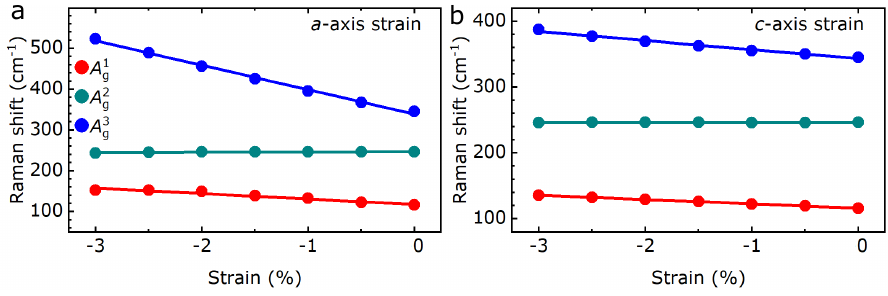}
    \caption{Raman shifts as a function of compressive strain along the (a) $a$-axis and (b) $c$-axis for the three $A_\textrm{g}$ modes of CrSBr. Here, the solid circles are the data points, while the lines are the linear fits.}
    \label{fig:strain}
\end{figure*}
%%%%%%%%%%%%%%%%%%%%%%%%%%%%%%%%%%%%%%%%%%%%%%%%%%%%%%%%%%%%%%%%%%%%%%%%%%%%%%%%%%%%%%%%%%%%
Although the two-host model reproduces the overall trends in the vibrational spectra, it implicitly assumes strain-independent interatomic force constants inherited from the parent compounds. As a consequence, changes in bond lengths and bonding strengths induced by halogen substitution are not explicitly captured. Consistent with this limitation, the calculated $A_\textrm{g}^3$ response develops a clear deviation from the experimental data, as discussed earlier.\par
To assess the specific role of lattice strain on the Raman-active phonons, we performed phonon calculations under controlled compressive deformation applied along the crystallographic $a$- and $c$-axes (figure \ref{fig:strain}). When compressive strain is applied along the $a$-axis, the $A_\textrm{g}^1$, $A_\textrm{g}^2$, and $A_\textrm{g}^3$ modes exhibit markedly different responses, with frequency shifts of 13.4 \cm/\%, -0.7 \cm/\%, and 60 \cm/\%, respectively. This pronounced sensitivity of the $A_\textrm{g}^3$ mode highlights its strong coupling to in-plane lattice distortions.\par
In case of compressive strain along the $c$-axis, the $A_\textrm{g}^2$ mode remains nearly unaffected, exhibiting a negligible slope of only 0.02 \cm/\%. The $A_\textrm{g}^1$ and $A_\textrm{g}^3$ modes, however, display clear hardening, with frequency shifts of 6.6 \cm/\% and 13.7 \cm/\%, respectively. These results demonstrate that the strain-induced modifications of bond lengths and force constants play a central role in determining the mode-dependent phonon shifts and provide a natural explanation for the residual discrepancies between the two-host model-based RGDOS-simulated and measured spectra.

%%%%%%%%%%%%%%%%%%%%%%%%%%%%%%%%%%%%%%%%%%%%%%%%%%%%%%%%%%%%%%%%%%%%%%%%%%%%%%%%%%%%%%

\subsection{Polarization-resolved Raman mapping and polarization switching}
\begin{figure*}[h]
    \centering
    \includegraphics[width=15cm]{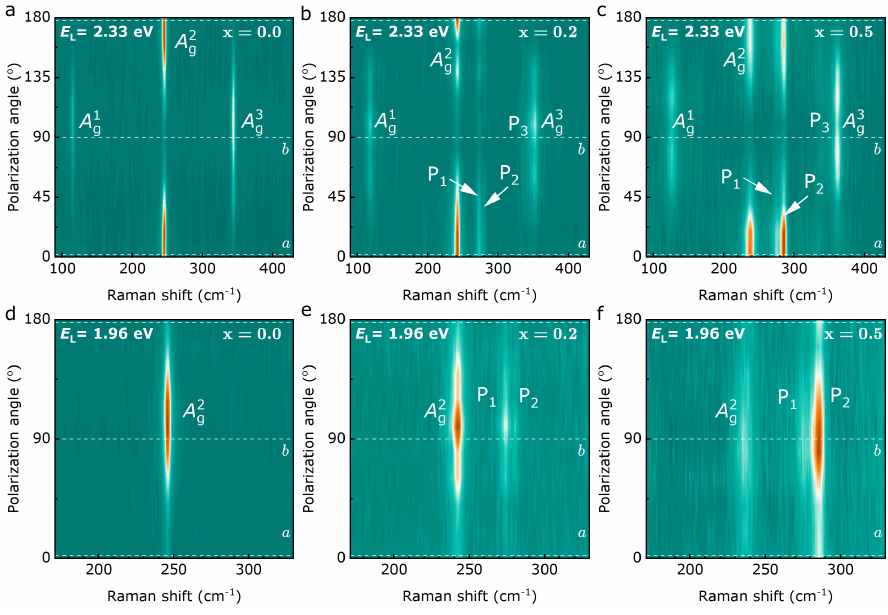}
    \caption{Polarization-resolved Raman maps for $x = 0.0, 0.2$, and 0.5 compositions for $E_\textrm{L}$ = 2.33 eV (a-c) and $E_\textrm{L}$ = 1.96 eV (d-f) excitation energies at 100 {\textmu}W power revealing the origin of the P modes and the characteristic $a$-axis to $b$-axis polarization switching of the $A_\textrm{g}^2$, P$_1$, and P$_2$ modes between off-resonance and on-resonance conditions.}
    \label{fig:Raman_maps}
\end{figure*}

%%%%%%%%%%%%%%%%%%%%%%%%%%%%%%%%%%%%%%%%%%%%%%%%%%%%%%%%%%%%%%%%%%%%%%%%%%%%%%%%%%%%%%%%%%%%%%
To better understand the origin of the additional P$_1$--P$_3$ spectral features, polarization-resolved Raman mapping was performed on the pristine CrSBr and the Cl-substituted samples under 2.33 eV excitation, as shown in figure \ref{fig:Raman_maps}a-c (for $x = 0.0, 0.2,$ and $0.5$) and supplementary information (Section S8; for $x = 0.3$ and $0.4$). The first-order $A_\textrm{g}$ modes maintain their characteristic \cite{Sahu2025, Mondal2025} angular dependencies, i.e., $A_\textrm{g}^1$ and $A_\textrm{g}^3$ remain maximized for E{$\parallel$}$b$, while $A_\textrm{g}^2$ is strongest for E{$\parallel$}$a$ indicating that the overall lattice symmetry is retained despite significant halogen substitution. Interestingly, the new modes follow polarization trends closely aligned with their nearest first-order counterparts: P$_1$ and P$_2$ display maxima along E{$\parallel$}$a$, akin to $A_\textrm{g}^2$, while P$_3$, observed for E{$\parallel$}$b$, resembles the angular behavior of $A_\textrm{g}^3$. This suggests that these features originate from modified local vibrational environments, with slightly shifted frequencies due to local distortions or electrostatic perturbations introduced by the random distribution of Cl atoms within the Br sublattice, rather than from the activation of silent Raman modes with different polarization symmetry.\par
To further explore the influence of excitation energy on the Raman response of Cl-substituted CrSBr, we conducted polarization-resolved Raman mapping under 1.96 eV excitation. As shown in figure \ref{fig:Raman_maps}d, for the pristine CrSBr sample ($x = 0$), the $A_\textrm{g}^2$ mode exhibits a distinct polarization switching. It reaches maximum when parallel to the $a$-axis under 2.33 eV excitation, which switches to maximum intensity when parallel to the $b$-axis under 1.96 eV excitation. This behavior reflects the excitation-energy-dependent modifications of the Raman tensor, which are governed by the anisotropic electronic structure of CrSBr and its coupling to specific phonon modes \cite{Mondal2025}.\par
Remarkably, in Cl-substituted samples (figures \ref{fig:Raman_maps}e–f), the same polarization switching is observed for the P$_1$ and P$_2$ modes. This confirms their shared vibrational origin with the $A_\textrm{g}^2$ mode and supports the presence of excitonic transitions near 1.96 eV that are active even in the presence of Cl substitution \cite{Klein2023, Sahu2025, Mondal2025, Qian2023}. The persistence of this switching behavior despite alloy disorder suggests that the essential features of the electronic band structure and its coupling to lattice vibrations remain intact. Furthermore, these newly emerging modes originate from modified local environments that remain strongly coupled to the electronic states, indicating robust anisotropic electron--phonon interactions.
%%%%%%%%%%%%%%%%%%%%%%%%%%%%%%%%%%%%%%%%%%%%%%%%%%%%%%%%%%%%%%%%%%%%%%%%%%%%%%%%%%%%%%%%%%%%%%%%%%
%%%%%%%%%%%%%%%%%%%%%%%%%%%%%%%%%%%%%%%%%%%%%%%%%%%%%%%%%%%%%%%%%%%%%%%
\subsection{High-order Raman features in Cl-substituted CrSBr}
%%%%%%%%%%%%%%%%%%%%%%%%%%%%%%%%%%%%%%%%%%%%%%%%%%%%%%%%%
\begin{figure*}[h]
    \centering
    \includegraphics[width=15cm]{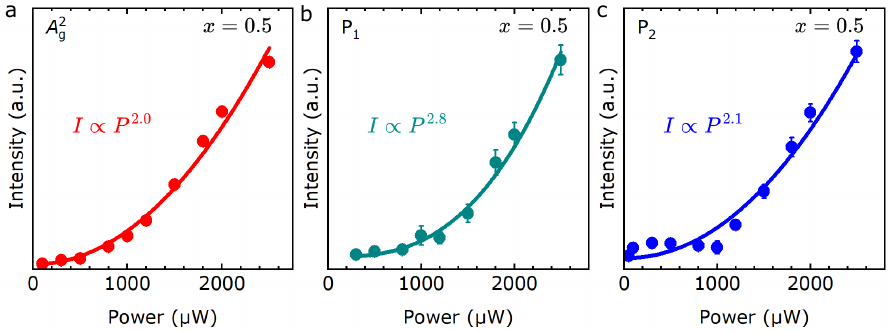}
    \caption{Evolution of the $A_\textrm{g}^2$ (a), P$_1$ (b), and P$_2$ (c) mode intensities as a function of laser power in the range of 100 {\textmu}W -- 2500 {\textmu}W, for $x = 0.5$ under $E_\textrm{L}$ = 1.96 eV excitation polarized parallel to crystalline $a$-axis.}
    \label{fig:SRS}
\end{figure*}
%%%%%%%%%%%%%%%%%%%%%%%%%%%%%%%%%%%%%%%%%%%%%%%%%%%%%%%%%%%%%%%%%%%%%%%%%%%%%%
Having established the polarization-resolved Raman behavior of Cl-substituted CrSBr under both non-resonant and resonant excitation conditions, we now turn to the nonlinear optical response of these alloys. In particular, we examine the manifestation of stimulated Raman scattering (SRS), a coherent light amplification process that emerges when the pump intensity exceeds a critical threshold, leading to directional amplification of specific phonon modes.\par
In our previous work \cite{Sahu2025}, pristine CrSBr exhibited a strong SRS-like feature under 1.96 eV excitation when the incident laser polarization was aligned along the $a$-axis, a condition that coincides with enhanced electron–phonon coupling due to excitonic resonance \cite{Klein2023, Qian2023}. This effect was marked by pronounced, intensity-dependent amplification and sharpening of the existing Raman modes, displaying clear signatures of a nonlinear behavior. Given that the Cl substitution alters both the vibrational landscape and the electronic environment, it becomes important to investigate whether such nonlinear behavior is preserved, suppressed, or possibly even enhanced in the alloyed system.\par
Figures \ref{fig:SRS}a--c, therefore, illustrate the evolution of the integrated Raman intensity ($I$) with increasing excitation power ($P$) for the $A_\textrm{g}^2$, P$_1$, and P$_2$ modes, respectively, in the $x = 0.5$ sample. For the $A_\textrm{g}^2$ mode (figure \ref{fig:SRS}a), a clear quadratic dependence $I \propto P^{2.0\pm0.1}$ is observed, consistent with our previous observation of SRS-like feature in pristine CrSBr \cite{Sahu2025}. This confirms that the $A_\textrm{g}^2$ mode remains the primary participant in the SRS-like process, even after Cl substitution. The onset of this nonlinear regime occurs beyond a threshold power of approximately 300 {\textmu}W (30 kW/cm$^{2}$), where the stimulated contribution begins to dominate the overall scattering intensity \cite{Sahu2025}. Notably, the P$_1$ and P$_2$ (figure \ref{fig:SRS}b-c) modes, which appear near the $A_\textrm{g}^2$ frequency and share similar polarization characteristics, also display supralinear power dependence with fitted exponents of $2.8\pm0.1$ and $2.1\pm0.1$, respectively. The emergence of SRS-like behavior in these modes, with comparable threshold, underscores their coupling to the same phonon excitation channels. This strong amplification confirms the presence of coherent phonon generation and further supports the interpretation that P$_1$ and P$_2$ are symmetry-perturbed counterparts of $A_\textrm{g}^2$ arising from the modified local vibrational environment due to Cl substitution.\par
Together, these results reveal that the SRS-like nonlinear response in CrSBr is maintained even upon Cl substitution, indicating that the underlying electron--phonon coupling remains strong and resilient across the alloy series. This robustness, combined with the material’s anisotropic optical properties and chemical tunability, opens avenues for developing polarization-sensitive tunable Raman lasers, nonlinear optical modulators, and integrated photonic devices based on van der Waals magnets.

%%%%%%%%%%%%%%%%%%%%%%%%%%%%%%%%%%%%%%%%%%%%%%%%%%%%%%%%%%%%%%%%%
\section{Conclusion}
In summary, this work demonstrates that halogen substitution in CrSBr provides a symmetry-selective pathway to phonon engineering that extends beyond simple mass-driven effects. By combining polarization-resolved Raman spectroscopy with first-principles lattice-dynamics modeling, we show that the alloy disorder activates symmetry-lowered phonon responses and simultaneously introduces anisotropic strain that selectively renormalizes specific vibrational modes. The interplay between configurational disorder and strain emerges as a key mechanism governing the evolution of the Raman spectrum in this low-symmetry van der Waals magnet. Importantly, the persistence of the polarization-dependent nonlinear Raman amplification under near-resonant excitation indicates that the strong electron–phonon coupling remains resilient despite substantial chemical substitution. These findings highlight the unique opportunities offered by low-symmetry layered materials, where controlled disorder can be used not merely as a perturbation but as a functional parameter to tailor vibrational and nonlinear optical properties.
%%%%%%%%%%%%%%%%%%%%%%%%%%%%%%%%%%%%%%%%%%%%%%%%%%

%\begin{suppinfo}

%\end{suppinfo}

%%%%%%%%%%%%%%%%%%%%%%%%%%%%%%%%%%%%%%%%%%%%%%%%%%%%%%%%%%%%%%%%%%%%%

\begin{acknowledgement}
S.S. and O.F. thank the Czech Science Foundation (project No. 23-06174K) for financial support. M.V. acknowledges the support of the Lumina Quaeruntur fellowship No. LQ200402201 by the Czech Academy of Sciences. Z.S. was supported by the ERC CZ program (project LL2101) from the Ministry of Education, Youth and Sports. A. H. and M. K. were supported by the Research Council of Finland (Flagship of Advanced Mathematics for Sensing Imaging and Modelling, grant 358944) and by Foundation PS. A.V.K. thanks the Deutsche Forschungsgemeinsachft (DFG, German Research Foundation) for support through the collaborative research center \say{Chemistry of Synthetic 2D Materials} SFB-1415-417590517 and Projects KR 4866/11-1. This work was also supported by the Ministry of Education, Youth, and Sports of the Czech Republic, Project No. CZ.02.01.01/00/22\_008/0004558, co-funded by the European Union. We also acknowledge the CzechNanoLab Research Infrastructure, supported by the Ministry of Education, Youth, and Sports of the Czech Republic (LM2023051). We also thank CSC-IT Center Science Ltd., Finland, for generous grants of computer time.
%\newline \textbf{Notes:} The authors declare no competing financial interest.
\newline \textbf{Data availability:} The data and analyses underlying this study are available from the HeyRACK repository at [persistent link to data repository TBC].
\newline \textbf{Supplementary information:} Data supporting our studies has been provided at the end of this document.
\end{acknowledgement}

\bibliography{bibliography}
%%%%%%%%%%%%%%%%%%%%%%%%%%%%%%%%%%%%%%%%%%%%%%%%%%%%%%%%%%%%%%%%%
\clearpage
\newpage
\section*{Graphical TOC}
\vspace{1cm}
\includegraphics[width=8.5cm]{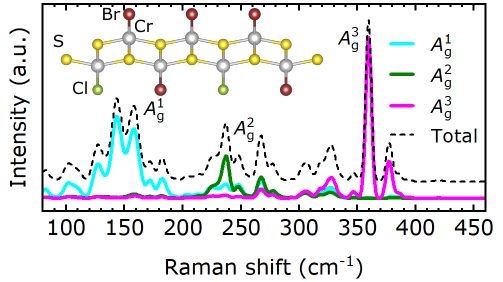}
Mode-resolved Raman of Cl-substituted CrSBr.

%%%%%%%%%%%%%%%%%%%%%%%%%%%%%%%%%%%%%%%%%%%%%%%%%%%%%%%%%%%%%
\end{document}